\documentclass[aps]{revtex4}
\usepackage[dvips]{graphicx}
\usepackage{amsmath}
\usepackage{latexsym}

\begin{document}

\title{String Approximation for Cooper Pair in 
High-T$_{\bf c}$ Superconductivity}
\author{V. Dzhunushaliev}
\email{dzhun@hotmail.kg}
\affiliation{Universit\"at Potsdam, Institute f\"ur Mathematik,
14469, Potsdam, Germany \\
and Phys. Dept., KRSU, 720000, Bishkek, Kyrgyzstan}

\vspace{1cm}

\begin{abstract}
It is assumed that in some sense the High-T$_c$ superconductivity 
is similar to the quantum chromodynamics (QCD). This means that 
the phonons in High-T$_c$ superconductor have the strong 
interaction between themselves like to gluons in the QCD. At the 
experimental level this means that in High-T$_c$ superconductor 
exists the nonlinear sound waves. It is possible that the 
existence of the strong phonon-phonon interaction leads 
to the confinement of phonons into a phonon tube (PT) 
stretched between two Cooper electrons like a hypothesized 
flux tube between quark and antiquark in the QCD. The flux tube 
in the QCD brings to a very strong interaction between quark-antiquark, 
the similar situation can be in the High-T$_c$ superconductor: 
the presence of the PT can essentially increase the binding 
energy for the Cooper pair. In the first rough approximation 
the PT can be approximated as a nonrelativistic string with Cooper 
electrons at the ends. The BCS theory with such potential 
term is considered. It is shown that Green's function 
method in the superconductivity theory is a realization of discussed 
Heisenberg idea proposed by him for the quantization of nonlinear 
spinor field. A possible experimental testing for the string 
approximation of the Cooper pair is offered.
\end{abstract}
\pacs{74.20.Mn  12.38.Lg}

\maketitle 

\section{Introduction}

In the quantum chromodynamics (QCD) there is a very fruitful 
analogy between superconductivity and QCD: according to the 't Hooft 
and Mandelstam assumption \cite{hooft}, \cite{mandel} 
a confinement in the QCD 
is similar to the dual Meissner effect in a superconductor. 
Such confinement scenario assumes that the ground state of QCD is a 
condensate of magnetic monopoles 
\footnote{we can name such condensate as a dual superconductor} which one 
confines (pinches) the color electric field of color charges 
into flux tubes 
\footnote{in ordinary superconductor this happens with magnetic 
field}. In other words, the magnetic charges, 
defined as the Dirac monopoles in ground state of QCD, 
should condense in the vacuum in just the same way as the 
Cooper pair in a superconductor. This analogy works in the direction: 
\textit{\textbf{superconductivity $\Rightarrow$ quantum chromodynamics.}} 
\par
An analogy can be offered in the opposite direction: 
from the QCD to the High-T$_c$ 
superconductivity by the following manner. An interaction between 
gluons exists in the QCD and it is so strong that the force lines 
of color gauge field are stretched into a flux tube between 
quark and antiquark. The most important in this picture 
is the strong interaction between gluons. 
This one distinguishes the quantum electrodynamics (QED) 
from the quantum chromodynamics: QED does not have the 
photon-photon interaction. Thus, our basic assumption is that 
\textit{\textbf{in the High-T$_c$ superconductor there is a strong 
phonon-phonon interaction}}. We assume that the jump from 
the ordinary superconductivity to High-T$_c$ 
one is analogous to the jump from QED to QCD.
\begin{center}
\fbox{%
\parbox{136pt}{\textbf{ordinary superconductivity} \\
\centerline {$\Downarrow$}\\
\textbf{High-T${\bf _c}$ superconductivity}}%
}
$\qquad \approx \qquad$
\fbox{%
\parbox{25pt}{\textbf{QED}\\
\centerline {$\Downarrow$}\\
\textbf{QCD}}%
}
\end{center}
The consequence of such conjectural analogy is that the 
interaction between phonons is so strong that 
\textit{the phonons between the Cooper pair are confined into a 
phonon tube (PT)} \cite{dzhun}. 
\par
The phonon Hamiltonian ${\cal H}_{ph}$ in the continuum limit we 
can write as follows
\begin{equation}
{\cal H}_{ph} = \int d^3{\bf r} 
\left [
\frac{\rho}{2} \dot s_i^2({\bf r},t) + 
c^{kl}\partial_ks_k \partial_ls^k + 
V({\bf r},t)
\right ] , 
\label{1-1}
\end{equation}
where $\rho$ is the mass density, ${\bf s} = \{s_i\}$  
($i=x,y,z$) is the ion deviation from the equilibrium state, 
$V$ is the potential energy, $c^{kl}$ are some coefficients. 
Our assumption indicates that the potential term $V$ 
in the High-T$_c$ superconductor 
can not be simplified as 
\begin{equation}
V \approx \frac{\partial ^2 V}{\partial s_i\partial s_j} 
s_i s_j
\label{1-2}
\end{equation}
even though for the small deviations $s_i$. 
The same takes place in the QCD with Lagrangian
\begin{equation}
{\cal L}_{QCD} \propto F^a_{\mu\nu}F^{a\mu\nu}
\label{1-3} , 
\end{equation}
here $\mu ,\nu = t,x,y,z$; 
$F^a_{\mu\nu} = \partial_\mu A_\nu - \partial_\nu A_\mu 
+ gf_{abc} A^b_\mu A^c_\nu$; $A^a_\mu$ is the SU(3) 
gauge potential; $f_{abc}$ are the structural constants of 
the SU(3) gauge group; $a,b,c$ are the color indexes. 
It is easy to see that in this QCD Lagrangian 
we have $(A)^3$ and $(A)^4$ terms. The most important is that 
coupling constant $g$ is not small, \textit{i.e.} in contrast 
with QED we have a very strong interaction between gluons that 
probably leads to the appearance of a nonlocal object - flux 
tube stretched between quark and antiquark. 

\section{BCS-theory + string approximation.}

The basic idea of BCS theory is that in the presence of even a 
weak interaction between electrons they cooperate in the 
Cooper pairs that leads to decreasing the ground state energy 
of superconductor in comparison with the normal state. According to  
the above-mentioned assumption about the possible analogy with the QCD 
we suppose 
\begin{itemize}
\item
There is a strong interaction between phonons (nonlinear potential
term in (\ref{1-1}) expression) which leads to appearing of a tube 
filled by phonons (like the flux tube filled by chromodynamical 
color fields in the QCD). We can name such tube as the PT.
\item
In the first approximation, neglecting by cross section 
of the PT, such object is like a string. Analogously to the QCD 
in the first rough approximation the PT in the High-T$_c$ 
superconductor can be considered as a string. 
\item
As in the QCD we assume that the interaction between Cooper 
electrons is described by the nonrelativistic string potential 
\begin{equation}
V = kl , 
\label{2-1}
\end{equation}
where $V$ is the potential energy of the phonon interaction 
between Cooper electrons, $k$ is a coefficient and $l$ is a length 
of the string.
\end{itemize}
\par 
In this model the Cooper pair is modeled as a string with 
two Cooper electrons attached at its ends. In the QCD such 
construction is a string with quark and antiquark at the ends. 
The interaction quark-antiquark is so strong that we have 
confinement: if we increase the string length then 
in some moment the string is torn and again we will have 
the quark-antiquark pair, \textit{i.e.} we can not obtain 
a single quark. By applying such model to the High-T$_c$ 
superconductivity we should be delicate: certainly a single 
electron exists 
\footnote{this means that the Cooper pair can be destroyed} 
but the basic idea is the same: the string interaction leads 
to an essential increase of the binding energy of the Cooper 
pair. 
\par
Now we would like to repeat the calculations in the microscopic 
BCS theory with the potential (\ref{2-1}) (following, for example, 
to Ref. \cite{tilley}). 
Let us assume that $|00\rangle$ is a quantum amplitude that the Cooper 
pair is unoccupied and $|11\rangle$ that it is occupied, then 
the wave function is 
\begin{eqnarray}
\Phi & = & \prod \limits_{\bf k} \psi _{\bf k} ,
\label{2-2}\\
\psi_{\bf k} & = & u^*_{\bf k} |00\rangle + 
v_{\bf k} |11\rangle , 
\label{2-3}
\end{eqnarray}
where $|u_{\bf k}|^2$ is the probability that the Cooper pair 
$({\bf k},\uparrow)$,  $(-{\bf k},\downarrow)$ is unoccupied 
and accordingly $|v_{\bf k}|^2$ is the probability that it is 
occupied; ${\bf k}$ is the wave vector. We have the normalization 
condition 
\begin{equation}
|v_{\bf k}|^2 + |u_{\bf k}|^2 = 1 .
\label{2-4}
\end{equation}
The normal ground state is given by 
\begin{eqnarray}
u_{\bf k} = 0 \quad & \mbox{and} & \quad 
v_{\bf k} = 1 \quad \mbox{for} \quad 
|{\bf k}| < {\bf k}_F , 
\label{2-5}\\
u_{\bf k} = 1 \quad & \mbox{and} & \quad 
v_{\bf k} = 0 \quad \mbox{for} \quad 
|{\bf k}| > {\bf k}_F , 
\label{2-6}
\end{eqnarray}
where ${\bf k}_F$ is the Fermi wave vector. 
\par
The energy of ground state is calculated by the variational method
\begin{equation}
\delta\langle \Phi |{\hat{\cal H}} - \mu {\hat N}|\Phi\rangle = 0
\label{2-7}
\end{equation}
with the constraint $\langle \Phi|{\hat N}|\Phi\rangle = N$, where 
${\hat N}$ is the number operator and $N$ 
is the mean number of particles in this system, 
$\mu$ is the Fermi energy. 
\par
The calculation for kinetic energy term ${\hat K}$ gives us 
\begin{equation}
\langle \Phi |{\hat K}|\Phi\rangle = 
2 \sum\limits_{\bf k} \varepsilon_{\bf k}|v_{\bf k}|^2 , 
\label{2-8}
\end{equation}
where the single particle energies $\varepsilon_{\bf k}$ are 
measured from the Fermi surface $\varepsilon_F$, the factor 
2 appears as $|v_{\bf k}|^2$ is the occupation probability for 
the Cooper pair. The expectation value of the potential energy 
part is 
\begin{equation}
\langle \Phi |{\hat V}|\Phi\rangle = \sum\limits_{\bf k, \bf k'} 
V_{\bf k \bf k'} u_{\bf k} v^*_{\bf k'}
u^*_{\bf k'} v_{\bf k} , 
\label{2-9}
\end{equation}
where $u^*_{\bf k'}v_{\bf k}$ is the amplitude for the 
initial state in which the Cooper pair ${\bf k}$ is occupied and 
pair ${\bf k'}$ in unoccupied; 
$u_{\bf k} v^*_{\bf k'}$ is the amplitude for the final state 
in which the reverse is true. Thus we have 
\begin{equation}
\langle \Phi |{\hat{\cal H}} - \mu {\hat N}|\Phi\rangle = 
2 \sum\limits_{\bf k} \varepsilon_{\bf k}|v_{\bf k}|^2 + 
\sum\limits_{\bf k, \bf k'} 
V_{\bf k \bf k'} u_{\bf k} v^*_{\bf k'}
u^*_{\bf k'} v_{\bf k} .
\label{2-10}
\end{equation}
As usual, we minimize this functional with respect to 
$v^*_{\bf k}$ that leads to the equation
\begin{equation}
\Delta_{\bf k} = - \sum\limits_{\bf k'} 
V_{\bf k \bf k'}\frac{\Delta_{\bf k'}}{2E_{\bf k'}}
\label{2-11}
\end{equation}
where 
\begin{eqnarray}
E^2_{\bf k} &= & \varepsilon^2_{\bf k} + \Delta^2_{\bf k}, 
\label{2-12}\\
\Delta_{\bf k} & = & \sum\limits_{\bf k} 
V_{\bf k \bf k'} u_{\bf k'} v_{\bf k'}
\label{2-13}
\end{eqnarray}
As usual we simplify
\begin{equation}
V_{\bf k \bf k'} = \left \{
\begin{array}{rl}
-V & \quad \mbox{if}\quad |\varepsilon_{\bf k}|,|\varepsilon_{\bf k'}| < 
\hbar \omega_D \\
0 & \quad \mbox{otherwise}
\end{array}
\right .
\label{2-14}
\end{equation}
here $\hbar \omega_D$ is the Debye energy of phonons. 
\par 
Now we can include our string assumption: we suppose that 
$V$ in the (\ref{2-14}) expression is 
\begin{equation}
V = kl , 
\label{2-15}
\end{equation}
where $k$ is some constant, $l$ is the length of Cooper pair 
(length of the string). 
\par 
As usual the further calculations give us
\begin{equation}
|\Delta| = \frac{\hbar \omega_D}{\displaystyle 
\sinh \left (\frac{1}{N_0V}\right )} , 
\label{2-16}
\end{equation}
where $N_0$ is the density of states in energy at 
the Fermi surface.

\section{Discussion}

We can hope that in the High-T$_c$ superconductor the quantity 
$V$ in the (\ref{2-16}) expression will be much more 
than the corresponding quantity in the ordinary superconductors 
\begin{equation}
V_{high} \gg V_{ordinary}
\label{3-1}
\end{equation}
as it takes place in the QCD: the strong interaction between 
quarks is much more than the electromagnetic one between 
electrons (positrons). In the QCD the interaction 
between gluons is so 
strong that the force lines are confined into the flux tube 
stretched between quark and antiquark.
\par
\begin{figure}[htb]
\begin{center}
\framebox[55mm]{
\includegraphics[height=5cm,width=5cm]{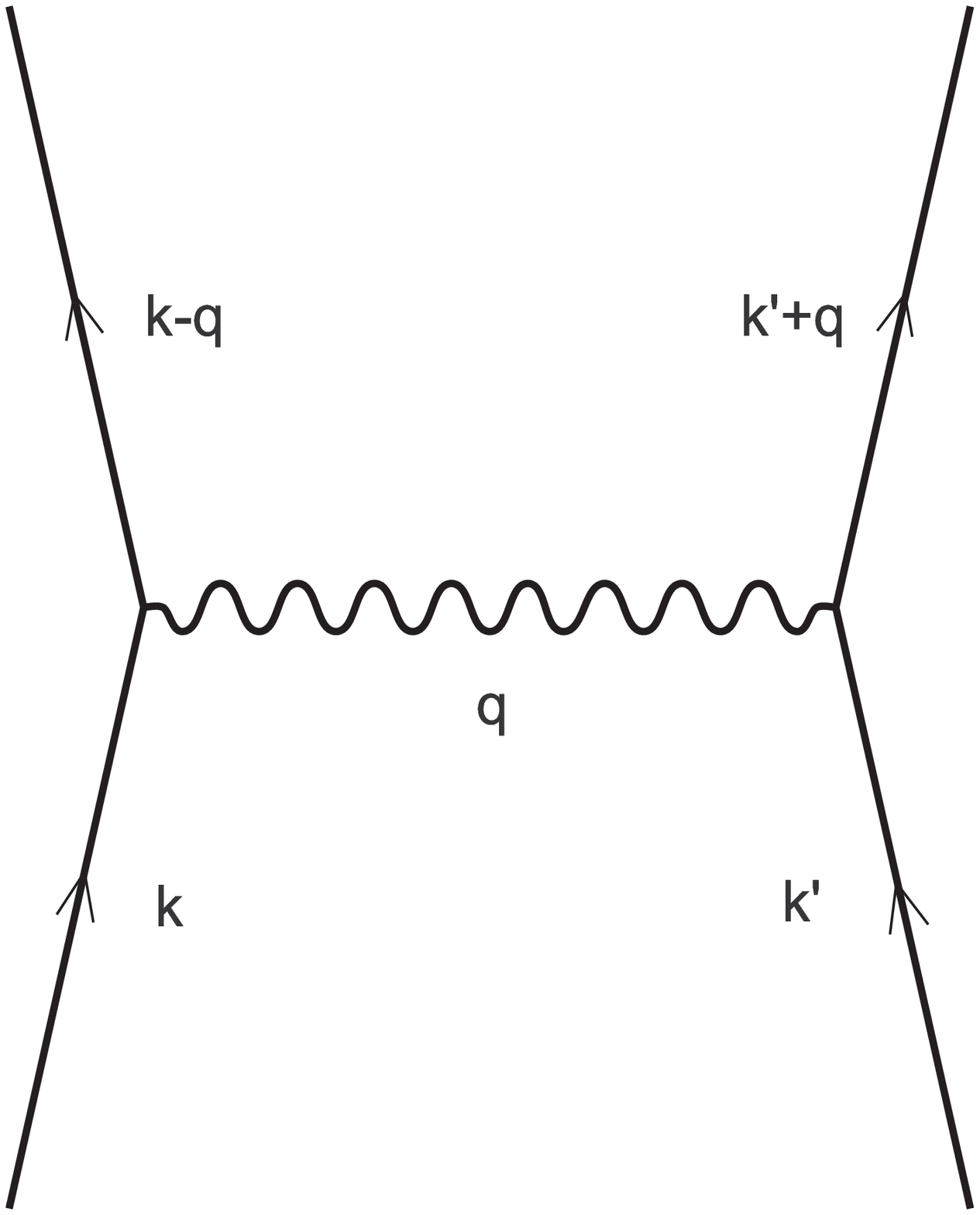}}
\caption{Phonon exchange between two electrons. 
Here the momenta of the incoming electrons 
are $\hbar {\bf k}$ and $\hbar {\bf k'}$ and the momentum of the 
exchanged phonon is $\hbar {\bf q}$. }
\label{fig1}
\end{center}
\end{figure}

The greatest difficulty here is the microscopical calculation 
of the quantity $V$. This difficulty is connected with the 
presence of the nonlinear potential term in the Hamiltonian. 
This means that in the presence of the strong self-interaction 
between phonons the expression for the exchange of one phonon between 
two electrons (see Fig. \ref{fig1}) 
\begin{equation}
V(\textbf{k,k}') = \frac{g^2\hbar \omega_{\bf q}}
{(\varepsilon _{\bf k+q})^2 - (\hbar \omega _{\bf q})^2}
\label{3-2}
\end{equation}
now is not correct. 
The same problem 
exists in the QCD: now we can not deduce the existence of the flux tube 
(string in the first rough approximation) from the SU(3) 
Lagrangian. This problem is connected with that we do not 
have appropriate mathematical tools in the nonperturbative QCD, 
\textit{i.e.} without the Feynman diagram techniques. 
\par
\begin{figure}[htb]
\begin{center}
\framebox[55mm]{
\includegraphics[height=5cm,width=5cm]{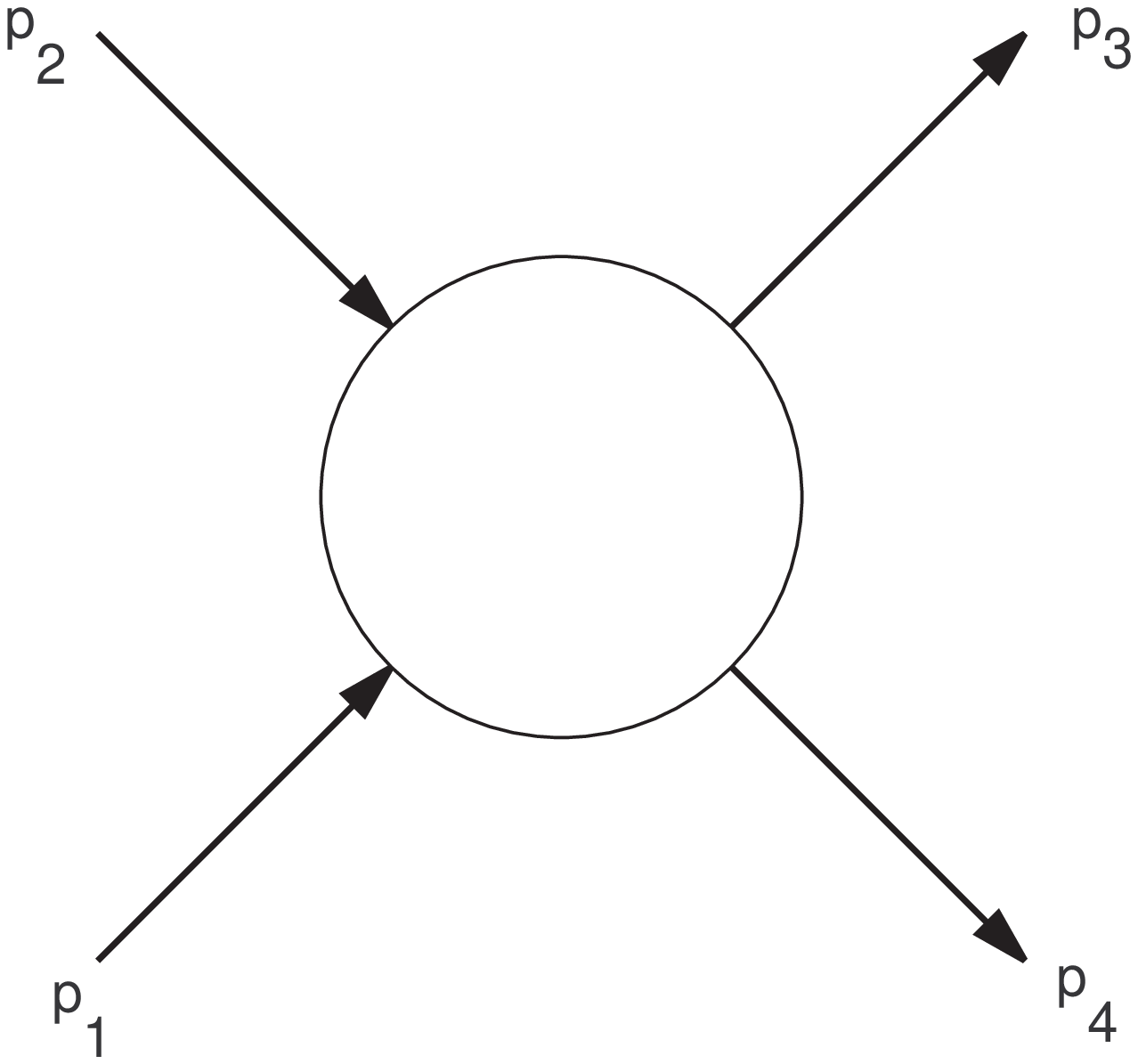}}
\caption{Elastic scattering of colliding particles with 
impulses $p_1$ and $p_2$ and divergent particles with 
impulses $p_3$ and $p_4$.}
\label{fig2}
\end{center}
\end{figure}

One of the first attempts of nonperturbative calculations 
in the QCD was a Veneziano amplitude \cite{venez} 
(see Fig.\ref{fig2}) 
\begin{equation}
A(s,t) = \frac{\Gamma\left (-\alpha(s)\right )
\Gamma\left (-\alpha(t)\right )}
{\Gamma\left (-\alpha(s) - \alpha(t)\right )}
\label{3-2a}
\end{equation}
where $s = -(p_1 + p_2)^2$ and $t = -(p_3 + p_4)^2$ are 
the Mandelstam variables; $\Gamma$ is the gamma function; 
$\alpha(s) = \alpha(0) + \alpha ' s$; $\alpha (0)$ and 
$\alpha '$ are some constants. 
This approach has not resulted in a great success in the QCD as 
this amplitude is some rough approximation for the correct 
4-point Green's function. It is possible to assume that such 
approximation can be made in the superconductivity theory : 
the scattering amplitude \eqref{3-2} of two electrons 
$V(\textbf{k,k}')$ 
is the Veneziano amplitude \eqref{3-2a}. Certainly (just as in QCD) 
it is a rough approximation for the correct Green's function. 
\par 
We can try to obtain an expression for $V$ on the 
basis of the dimensional reasons. The result, for example, can be 
\begin{eqnarray}
k & \approx & \frac{\hbar \omega_D}{\sqrt S} ,
\label{3-3}\\
V & \approx & \hbar \omega_D\frac{l}{\sqrt S} , 
\label{3-4}
\end{eqnarray}
where $S$ is the cross section of the PT. If we suppose that the PT 
is located inside of a layer of an anisotropical superconductor then 
$\sqrt S \approx \delta$, where $\delta$ is the thickness of  
the layer. In this case these expressions will be
\begin{eqnarray}
k \approx \frac{\hbar \omega_D}{\delta} , 
\label{3-5}\\
V \approx \hbar \omega_D\frac{l}{\delta}
\label{3-6}
\end{eqnarray}
because of $l \gg \delta$ we have $V \gg \hbar \omega_D$. 

\section{Phonon tube}

What is the physical meaning of the PT ? In the QCD the flux tube is 
filled with the color field which is parallel to the tube axis and equal 
to zero outside this tube. In our case we can suppose that 
there can be one from the following possibilities
\begin{alignat}{3}
\langle {\bf s}\rangle   & \neq 0 \quad \text{and/or} & \qquad
\langle {\bf s}^2\rangle & \neq 0 \quad \text{and/or} & \qquad
\langle {\bf s}^3\rangle & \neq 0 \quad \ldots \quad 
\text{inside PT and}
\label{4-1}\\
\langle {\bf s}\rangle   & \approx 0 \quad \text{and} & \qquad
\langle {\bf s}^2\rangle & \approx 0 \quad \text{and} & \qquad
\langle {\bf s}^3\rangle & \approx 0 \quad \ldots \quad 
\text{outside PT} .
\label{4-2}
\end{alignat}
The first possibility $\langle {\bf s}\rangle \neq 0$ 
means that the nonzero static deviation of the ion lattice 
exists only inside PT. The second one 
$\langle {\bf s}^2\rangle \neq 0$ 
means that the nonzero mean square of the 
deviation exists only inside PT only and so on. 

\section{Heisenberg quantization model for the PT}

A possible strong phonon-phonon interaction can
obstruct the application of Feynman diagram techniques in our
case. Some time ago W. Heisenberg had conceived of the difficulties
in applying an expansion in small parameters to quantum
field theories having strong interactions. He had investigated
the Dirac equation with nonlinear terms (Heisenberg equation)
(see, for example, Ref's \cite{heis1} - \cite{heis2}).
In these papers he repeatedly underscored that a nonlinear
theory with a large parameter 
\textit{\textbf{requires the introduction of another
quantization rule.}} He worked out a quantization method for
strong nonlinear field unusing the expansion in a small parameter. 
It is possible that in the High - T$_c$ 
superconductivity the interaction between phonons is strong making
it necessary take into account the interaction between
phonons to correctly calculate the energy of the Cooper pairs.

\subsection{Heisenberg quantization of nonlinear spinor field}
\label{a}

Heisenberg's basic idea proceeds from the fact that the n-point Green
functions must be found from some infinity differential equations system
derived from the field equation for the field operator. For example,
we present Heisenberg quantization for nonlinear spinor field.
\par
The basic equation (Heisenberg equation) has the following
form:
\begin{equation}
\gamma ^\mu \partial _\mu  \hat\psi (x) -
l^2 \Im \left [ \hat\psi (\hat{\bar\psi}  \hat\psi)\right ] = 0 ,
\label{a1}
\end{equation}
where $\gamma ^\mu $ are Dirac matrices; $ \hat\psi ({\bf x})$ is the field
operator; $\hat{\bar\psi}$ is the Dirac adjoint spinor;
$\Im [\hat\psi (\hat{\bar\psi}\hat\psi )]=  \hat\psi (\hat{\bar\psi} 
\hat\psi)$ or $ \hat\psi \gamma ^5(\hat{\bar\psi} \gamma ^5 \hat\psi)$ or
$\hat\psi \gamma ^\mu (\hat{\bar\psi} \gamma _\mu \hat\psi)$ or
$\hat\psi \gamma ^\mu \gamma ^5(\hat{\bar\psi} \gamma _\mu \gamma ^5
\hat\psi)$. Heisenberg emphasizes that the 2-point Green function
$G_2({\bf x}_2,{\bf x}_1)$ in this theory differs strongly  
from the propagator
in linear theory. This difference lies in its behaviour on
the light cone.
$G_2({\bf x}_2,{\bf x}_1)$ oscillates strongly on the light cone in contrast
to the propagator of the linear theory which has a $\delta$-like
singularity. Then Heisenberg introduces the $\tau$ functions:
\begin{equation}
\tau ({\bf x}_1{\bf x}_2\ldots |{\bf y}_1{\bf y}_2\ldots ) =
\langle 0|T\psi ({\bf x}_1)\psi ({\bf x}_2)\ldots \psi ^*({\bf y}_1)
\psi ^*({\bf y}_2) \ldots |\Phi \rangle,
\label{a2}
\end{equation}
where $T$ is the time ordering operator. $|\Phi\rangle$ is a system
state characterized by the fundamental Eq. (\ref{a1}). Relationship
(\ref{a2}) allows us to establish a one-to-one
correspondence between the system state $|\Phi\rangle$ and
the function set $\tau$. This state can be defined using the infinite
function set of (\ref{a2}). Applying Heisenberg's equation (\ref{a1}) to
(\ref{a2}) we can obtain the following infinite equations
system:
\begin{eqnarray}
l^{-2} \gamma ^\mu _{(r)}\frac {\partial}{\partial x^\mu _{(r)}}
\tau ({\bf x}_1\ldots {\bf x}_n|{\bf y}_1\ldots {\bf y}_n) =
\Im \left [\tau ({\bf x}_1\ldots {\bf x}_n {\bf x}_r|
{\bf y}_1\ldots {\bf y}_n {\bf y}_r)\right ] +
\nonumber \\
\delta ({\bf x}_r - {\bf y}_1) \tau 
\left ({\bf x}_1\ldots {\bf x}_{r-1}{\bf x}_{r+1}\ldots {\bf x}_n|
{\bf y}_2\ldots {\bf y}_{r-1}{\bf y}_{r+1}\ldots {\bf y}_n\right ) +
\nonumber \\
\delta ({\bf x}_r - {\bf y}_2) \tau \left ({\bf x}_1
\ldots {\bf x}_{r-1}{\bf x}_{r+1}\ldots {\bf x}_n|
{\bf y}_1{\bf y}_2\ldots {\bf y}_{r-1}
{\bf y}_{r+1}\ldots {\bf y}_n\right ) + \ldots .
\label{a3}
\end{eqnarray}
Heisenberg then employs the Tamm - Dankoff method for getting approximate
solutions to the infinite equations system of (\ref{a3}). The key to this
method lies in the fact that the system of equation has an approximate
solution derived after cutting off the infinite equation system
(\ref{a3}) to a finite equation system.
\par
It is necessary to note that a method of solution to Eq. (\ref{a3})
can be various. For example, we can try to determine
the Green's functions using the numerical lattice calculations.
Here the important point is the following: The technique of expansion
in small parameters (Feynman diagrams) can not be employed
for strong nonlinear fields. It is possible that as in quantum
chromodynamics, where quarks are thought to interact strongly 
by means of flux tubes, so too
in High-T$_c$ superconductivity phonons may strongly interact
among themselves.
\par
In Ref. \cite{vds,dzhun2} such a mechanism is applied for the QCD. 
In this paper we apply a variant of Heisenberg's quantization method 
to solutions of the classical SU(3) Yang-Mills 
field equations which have bad asymptotic behavior. 
After quantization it has been found that the bad 
features ({\it i.e.} divergent 
fields and energy densities) of these solutions 
are moderated. From these results is argued that in 
general the n-point Green's functions for Yang-Mills 
theories can have nonperturbative pieces which 
can not be represented as the sum of Feynman diagrams.  
\par
In Heisenberg's theory the matter and the interacting fields
are identical: fundamental spinor field $\psi (x)$. From a more
recent perspective this is not the case. An interaction is
carried by some kind of boson field. In superconductivity this
is the phonons, in quantum chromodynamics it is the nonabelian $SU(3)$ 
gauge field - gluons.
\par
In conclusion of this section we emphasize again Heisenberg's
statement that 
\textit{\textbf{the perturbation theory is inapplicable
to strong nonlinear fields.}} 

\subsection {Heisenberg quantization method for High-T$_c$
superconductivity}

Thus, the basic assumption supposed here is the following: 
\textbf{\textit{The energy of Cooper pair has an essential
contribution coming from an interaction of phonons}}. This
means that the corresponding sound wave is a nonlinear wave. 

\subsubsection{An application of Heisenberg quantization 
method for the Green's function method.}

In this subsection we would like to show that 
Heisenberg idea about quantization of strongly interacting 
fields in fact had been applied in the superconductivity 
theory for definition of Green's function. 
Here we would like to show that the calculations which 
was made in Ref.~\cite{gor} is an application of the 
Heisenberg quantization idea. 
In this section we follow to Ref.\cite{abr}. 
\par 
The Hamiltonian of the system electrons describing the 
properties of a metal in the superconductivity state is 
\begin{equation}
    \hat{H} = \int
    \left [
        -\left (
        \hat\psi^+_\alpha \frac{\nabla^2}{2m}\hat\psi_\alpha
         \right ) + \frac{\lambda}{2}
         \left (
         \hat\psi^+_\beta \left (
                          \hat\psi^+_\alpha \hat\psi_\alpha
                          \right )
         \hat\psi_\beta
         \right )
    \right ] d V ,
\label{green1}
\end{equation}
where $\hat\psi_\alpha$ is the operator of spinor field describing 
electrons, $m$ is the electron mass, $\lambda$ is some constant 
and $\alpha , \beta$ are the spinor indexes. As usually 
(it is the same as Heisenberg equation \eqref{a1} for 
nonlinear spinor field) 
the operators $\hat\psi$ and $\hat\psi^+$ obey the following 
operator equations 
\begin{eqnarray}
\left (
i\frac{\partial}{\partial t} + \frac{\nabla ^2}{2m} 
\right )
\hat\psi_\alpha (x) - \lambda 
\left (
\hat\psi^+_\beta (x) \hat\psi_\beta (x)
\right )\hat\psi_\alpha (x) & = & 0 ,
\label{green2}\\
\left (
i\frac{\partial}{\partial t} - \frac{\nabla ^2}{2m} 
\right )
\hat\psi^+_\alpha (x) + \lambda 
\hat\psi^+_\alpha (x) 
\left (
\hat\psi^+_\beta (x)
\hat\psi_\beta (x) 
\right )& = & 0 .
\label{green3}
\end{eqnarray}
As well as in Heisenberg method for nonlinear spinor field 
we have an equation for the 2-point Green's function 
$G_{\alpha\beta}(x,x') = -i \langle T(\hat\psi_\alpha (x)
\hat\psi^+_\beta (x'))\rangle$ 
\begin{equation}
\left (
i\frac{\partial}{\partial t} + \frac{\nabla ^2}{2m} 
\right )G_{\alpha\beta}(x,x') + 
i\lambda  \langle T
\left (
\hat\psi^+_\gamma (x)\hat\psi_\gamma (x) 
\hat\psi_\alpha (x) \hat\psi^+_\beta (x') 
\right )\rangle = \delta (x - x') .
\label{green4}
\end{equation}
Further we have to write an equation for term 
$\langle T (\hat\psi^+_\gamma (x)\hat\psi_\gamma (x) 
\hat\psi_\alpha (x) \hat\psi^+_\beta (x') )\rangle$ and 
so on. After this we will have an infinite set for Green's 
function. The main problem here is: how can we cut off this 
infinite equation system ? Let us cite Ref.\cite{abr}: 
``$\ldots$ For interacting particles, 
the product of four $\hat\psi$-operators can be expressed 
in terms of the vertex part, \textit{i.e.} it already 
includes the contributions from various scattering processes. 
In the weak-interaction model under consideration, 
these scattering processes involving collisions of particles 
can be neglected, but at the same time, it must be borne 
in mind that the ground state of the system differs from 
the usual state with a filled Fermi sphere, because of 
the presence of bound pairs of electrons. $\ldots$''. 
This means that 
in the textbook \cite{abr} it was made the following 
approximation: the operators $\hat\psi \hat\psi$ and 
$\hat\psi^+ \hat\psi^+$ contain terms corresponding to the 
annihilation and creation of bound pairs (Cooper pairs). It 
allows to state 
\begin{equation}
\begin{split}
\left \langle T \left (
\hat\psi_\alpha (x_1) \hat\psi_\beta (x_2) 
\hat\psi^+_\gamma (x_3) \hat\psi^+_\delta (x_4) 
                \right ) 
\right \rangle
\approx \\ 
-\left \langle T \left (
\hat\psi_\alpha (x_1) \hat\psi^+_\gamma (x_3) 
                 \right )
\right \rangle 
\left \langle T \left (
\hat\psi_\beta (x_2) \hat\psi^+_\delta (x_4) 
                 \right )
\right \rangle  + \\
\left \langle T \left (
\hat\psi_\alpha (x_1) \hat\psi^+_\delta (x_4) 
                 \right )
\right \rangle 
\left \langle T \left (
\hat\psi_\beta (x_2) \hat\psi^+_\gamma (x_3) 
                 \right )
\right \rangle  + \\
\left \langle N \left\vert 
T \left (
\hat\psi_\alpha (x_1) \hat\psi_\beta (x_2) 
  \right ) \right\vert N + 2
\right \rangle
\left \langle N +2 \left\vert
T \left (
\hat\psi^+_\gamma (x_3) \hat\psi^+_\delta (x_4) 
  \right ) \right\vert N 
\right \rangle
\end{split}
\label{green5}
\end{equation}
where $|N \rangle$ and $|N + 2 \rangle$ are ground states 
of system with $N$ and $N +2$ particles (Cooper pairs), 
respectively. We should note that this expression is 
approximate one and following to Heisenberg idea 
it allows us to cut off the above-mentioned 
infinite equation set for Green's function. 
The key role here plays the last term in expression 
\eqref{green5}. This expression allows us to split the 
4-point Green's function with the nonperturbative way 
not using Feynman diagram techniques.
\par 
Eq. \eqref{green5} means that we have neglected 
all effects of scattering particles by each other 
and the presence of the interaction has been taken into 
account only as leading to the formation 
of bound pairs. The third term in the right-hand side of Eq.  
\eqref{green5} has been written in complete analogy with 
the case of a Bose gas in the correspondence with the 
fact that all bound pairs (Cooper pairs) 
are ``condensed on the lowest level''. 
The quantity 
\begin{equation}
\left \langle N \left\vert 
T \left (
\hat\psi_\alpha \hat\psi_\beta 
  \right ) \right\vert N + 2
\right \rangle
\left \langle N +2 \left\vert 
T \left (
\hat\psi^+_\gamma \hat\psi^+_\delta 
  \right ) \right\vert N 
\right \rangle ,
\label{green6}
\end{equation}
obviously has the same order of magnitude as the density 
of pairs. 
We can introduce the following functions 
\begin{eqnarray}
e^{-2i\mu t} F_{\alpha\beta}(x - x') & = & 
\left \langle N \left\vert 
T \left (
\hat\psi_\alpha (x) \hat\psi_\beta (x') 
  \right ) \right\vert N + 2
\right \rangle
\label{green6a} \\
e^{2i\mu t} F^+_{\alpha\beta} (x - x') & = & 
\left \langle N +2 \left\vert 
T \left (
\hat\psi^+_\gamma (x) \hat\psi^+_\delta (x') 
  \right ) \right\vert N 
\right \rangle ,
\label{green6b}
\end{eqnarray}
where $\mu$ is a chemical potential. 
We now substitute \eqref{green5} into the equation 
\eqref{green2} for the Green's function. 
We can everywhere omit the first two terms in the right-hand
side of \eqref{green5} since they 
lead to an additive correction to the chemical potential 
in the equations for the functions $G, F, F^+$. 
As a result we obtain the following equation 
connecting $G$ and $F^+$:
\begin{equation}
\left (
i\frac{\partial}{\partial t} + \frac{\nabla ^2}{2m} 
\right )G(x - x') - i\lambda F(0+) 
F^+(x - x') = \delta(x - x') .
\label{green7}
\end{equation}
The quantity $F(0+)$ is defined as 
\begin{equation}
F_{\alpha\beta}(0+) = 
e^{2i\mu t}
\left \langle 
N \left\vert \hat\psi_\alpha(x) \hat\psi_\beta(x) \right\vert N +2 
\right \rangle = 
\underset{
\substack{
x \rightarrow x', \\ t \rightarrow t' + 0
          }
}
{\lim} F_{\alpha\beta} (x - x') .
\label{green8}
\end{equation}
An equation for $F^+(x - x')$ can be obtained in a similar way 
by using the equation \eqref{green3} 
\begin{equation}
\left (
i\frac{\partial}{\partial t} - \frac{\nabla ^2}{2m} - 2\mu
\right )F^+(x - x') + i\lambda F^+(0+) 
G(x - x') = 0 .
\label{green9}
\end{equation}
Analogously to Eq.\eqref{green8} we have 
\begin{equation}
F^+_{\alpha\beta}(0+) = 
e^{-2i\mu t}
\left \langle 
N+2\left\vert \psi^+_\alpha(x) \psi^+_\beta(x) \right\vert N
\right \rangle 
\label{green10}
\end{equation}
Now we have Eq's.\eqref{green7} and \eqref{green9} as 
the finite set of equations 
for the 2-point Green's functions $G$ and $F$.
\par 
Thus, above-mentioned reasonings should convince 
us that \textit{\textbf{this Green's function method 
in the superconductivity 
theory is a realization of Heisenberg idea developed 
by him for the quantization of non-linear spinor field.}} 

\subsubsection{Phonon-phonon interaction}

In this subsection we would like to consider a possibility 
for the phonons to form a static configuration. 
Let us write again the Lagrangian (\ref{1-1}) 
for phonons in continuous limit 
\begin{equation}
{\cal L}_{ph} = \int d^3{\bf r} 
\left [
\frac{\rho}{2} {\dot s}_i^2({\bf r},t) + 
c^{kl}\partial_ks_k \partial_ls^k - 
V({\bf r},t)
\right ] .
\label{d1}
\end{equation}
Thus, in this model it is assumed that operators of strong
nonlinear fiels $s_i$ must satisfy the following
equation (which is implied from the Lagrangian (\ref{d1})):
\begin{equation}
\frac{\partial^2 \hat s_i}{\partial t^2} + 
c^{kl}\frac{\partial^2 \hat s_i}{\partial x^k \partial x^l} 
= - \frac{d V(\hat s_i)}{d \hat s_i} .
\label{d2}
\end{equation}
The multitime formalism of Heisenberg's method (when in
$\tau (t_1, t_2, \cdots ), t_1 \ne t_2 \ne \cdots $)
allows us to investigate the scattering processes in quantum
theory. The simultaneous formalism (when
$\tau (t_1, t_2, \cdots ), t_1 = t_2 = \cdots =t$)
allows us to calculate the mean value
of the field, the energy, or any combination of field powers.
\par
\begin{figure}[htb]
\begin{center}
\framebox[55mm]{
\includegraphics[height=5cm,width=5cm]{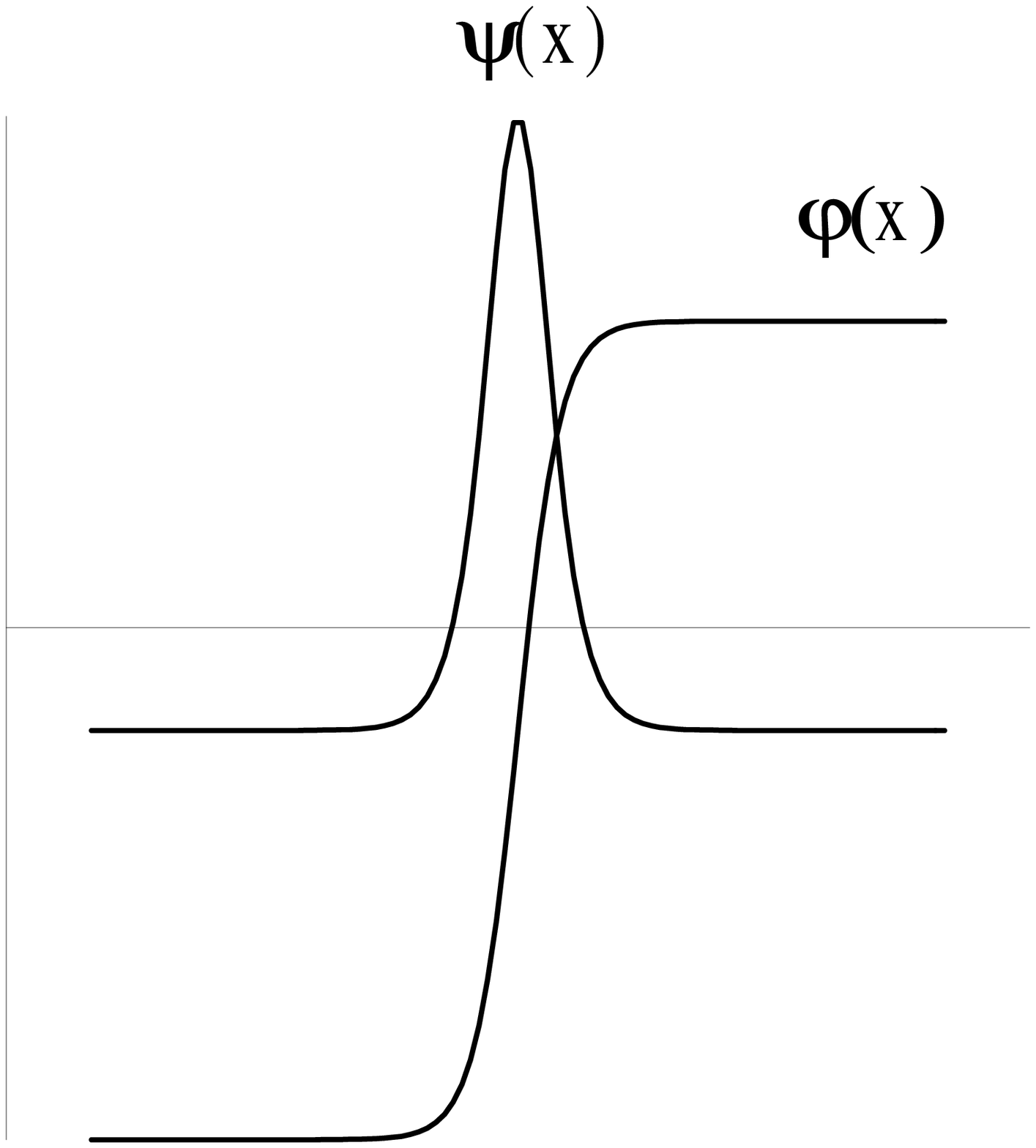}}
\caption{Zero energy fermion in a soliton background.}
\label{fig3}
\end{center}
\end{figure}

 
Let us consider, for example, the simplest case of the 
scalar field when the vector $\bf s$ is replaced by the 
scalar field $\varphi$. In this case the Lagrangian is 
\begin{equation}
{\cal L} = \int d^3 {\bf r} \left [
\left (\partial _\mu\varphi \right )^2 - V(\varphi) 
\right ] 
\label{d3-1}
\end{equation}
where $\mu = t,x,y,z$ and the potential term 
\begin{equation}
V(\varphi )= \frac{\lambda}{4}\left (
\varphi ^2 - \varphi_0^2 
\right ) .
\label{d3-1a}
\end{equation}
The classical field equation is
\begin{equation}
\Box \varphi (\bf r)= - \lambda \varphi (\bf r)
\left(
\varphi ^2(\bf r) - \varphi_0^2 
\right) .
\label{d3-2}
\end{equation}
The quantization of (\ref{d3-2}) equation gives us
\begin{equation}
\Box \hat \varphi ({\bf r}) = \widehat {\varphi ^3({\bf r})} -
\hat \varphi ({\bf r})\varphi _0^2 .
\label{d3-3}
\end{equation}
It is easy to see that the mean value
$\langle \varphi ({\bf r})\rangle  = 
\langle 0|\hat \varphi ({\bf r}) |0\rangle $
satisfies the following equation:
\begin{equation}
\Box \langle \varphi ({\bf r})\rangle  = 
\langle \varphi ^3({\bf r})\rangle  -
\varphi _0^2\langle \varphi ({\bf r})\rangle .
\label{d3}
\end{equation}
For the definition of $\langle \varphi ^3({\bf r})\rangle $ 
we turn to Eq.(\ref{d3-3}) and obtain 
($\langle \varphi ^3({\bf r})\rangle  = \tau ({\bf r}{\bf r}{\bf r}$) 
in Heisenberg's notation):
\begin{equation}
\Box \langle \varphi ^3({\bf r})\rangle  = 3
\lambda \left (
\langle \varphi ^5({\bf r})\rangle  - 
\varphi _0^2\langle \varphi ^3({\bf r})\rangle 
\right ),
\label{d4}
\end{equation}
here $\langle \varphi ^5({\bf r})\rangle  = 
\tau ({\bf r}{\bf r}{\bf r}{\bf r}{\bf r})$.
Analogously it can be used to derive the infinite equation 
system for calculating $\langle \varphi ^n({\bf r})\rangle $. 
In the first approximation we can solve
this equation system using the following  assumption:
\begin{equation}
\langle \varphi ^3({\bf r})\rangle  \approx \langle \varphi 
({\bf r})\rangle ^3,
\label{d5}
\end{equation}
and then we can derive the equation 
\begin{equation}
\Box\langle\varphi ({\bf r})\rangle = 
\lambda \langle\varphi ({\bf r})\rangle 
\left (
\langle\varphi ({\bf r})\rangle ^2 - \varphi_0^2
\right ) .
\label{d5-1}
\end{equation}
This equation is very interesting for us. Derrick's Theorem 
\cite{derr} states that Eq. \eqref{d5-1} has a static solution with 
the finite energy only in two dimensional spacetime $(t,x)$. 
This solution is 
\begin{equation}
\langle \varphi (x) \rangle = \varphi _1 \tanh 
\left [
\frac{1}{2}m\left (x - x_0 \right )
\right ] 
\label{d5-2}
\end{equation}
where $\varphi_1, m$ and $x_0$ are some constants. 
The first assumption is that such solution can be realized on the 
layer of High-T$_c$ superconductor as a line (wall) separating 
regions with different vacua ($\varphi = \pm \varphi _0$). 
The existence of regions with different vacua can have very 
interesting physical consequences for a distribution of free 
electrons in the ion lattice. Following Ref.\cite{jackiw} 
we present a model example of a fermion coupled to solution 
\eqref{d5-2}. Let we have 2-dimensional Dirac equation 
\begin{equation}
\left (
\gamma^\mu \partial_\mu + g \varphi (x)
\right ) \psi (x) = 0
\label{d5-3}
\end{equation}
where $\psi = \tbinom{\psi_1}{\psi_2}$ is the two-component spinor; 
$g$ is the coupling constant; $\varphi (x)$ is the soliton 
\eqref{d5-2}; $\gamma^\mu$ are the Dirac matrices, which we can 
choose to be the Pauli matrices 
\begin{equation}
\gamma^1 = \sigma _1, \; \gamma^4 = \sigma _3 ,
\end{equation}
Now we search a static solution of Dirac equation in the presence 
of the soliton 
\begin{equation}
\left (
\sigma _1\partial _x + g \varphi
\right )\psi = E \sigma _3 \psi = 0
\label{d5-4}
\end{equation}
here we consider case with energy $E = 0$. In this case 
we have the following normalizable zero mode 
\begin{equation}
\psi = \psi_0 
\left \{
    \cosh 
    \left [ 
    \frac{m}{2}\left ( x - x_0 \right )
    \right ]
\right \}^{-2m/m_\psi}
\dbinom{1}{1}
\label{d5-5}
\end{equation}
where $m_\psi = g\varphi_0$. 
We see that it is strongly localized at the position $x_0$ 
of the soliton (see, Fig.\ref{fig3}). In the context of 
mechanism with different vacua this toy model can mean 
that the free electrons on the layer of the High-T$_c$ superconductor 
will be located on the boundaries between regions 
with different vacua. 
\par 
Evidently, the appearance of localized solutions is connected 
with the presence of two vacuums ($\varphi = \pm \varphi _0$). 
It allows us 
to assume that the mechanism offered here for the formation 
of Cooper pair in High-$T_c$ superconductivity depends on the 
existence of several vacuums for phonon field, \textit{i.e.} 
on the presence of several equilibrium states (stable and 
unstable) for ions of the lattice. 
\par
Such interpretation 
depends on the existence of different vacua. Thus, this mechanism 
takes place only in the presence of different equilibrium states 
for ions of the lattice. It is necessary to note that can exist other 
mechanism of the field localization in some region. It is possible, 
for example, that such mechanism (without different vacua of gauge 
field) occurs in QCD by forming a hypothesized flux tube 
between quark-antiquark. 
\par 
Another possibility can be connected with cut-in of an external 
gauge field. In the Ref. \cite{nielsen} it was shown 
that the scalar field coupled with the electromagnetic field 
can forms the flux tube (Nielsen - Olesen flux tube). The 
detailed investigation of such possibility in the 
context of the Gingbug-Landau theory is given 
in Ref.\cite{obukh}. 
\par
It should be pointed out that the investigation of
$\tau ({\bf r}{\bf r}\cdots ) = \langle \varphi ^n({\bf r})\rangle $
gives us the information about the mean value of the field $\varphi ({\bf r})$.
For the investigation of questions on the scattering or interaction
of phonons it is necessary to explore the functions,
$\tau ({\bf r}_1{\bf r}_2\cdots ) = 
\langle 0|\varphi ({\bf r}_1)\cdots \varphi ({\bf r}_n)|0\rangle $. 

\section{An experimental testing}

How can this string model for the Cooper pair in 
the High-T$_c$ superconductor be proved ? The best way is the direct 
observation of the PT between Cooper electrons. Evidently it 
should be connected with the detection of the ion lattice between 
Cooper electrons. If we can establish that 
$\langle s^n_i \rangle \neq 0$ inside of the tube and 
$\langle s^n_i \rangle \approx 0$ outside of the PT then 
probably it will indicate the existence of the PT. 
Certainly, such measurements of the state of ion lattice 
are very difficult. 
\par
Another indirect way is a test for the nonlinear potential 
$V({\bf r})$. The presence of such nonlinear term can lead 
to: (a) the nonlinear effects by propagation of the sound wave 
in the High~-~T$_c$ superconductor along the layer; (b) 
the different equilibrium states of ions in the lattice.
Such effects can be: (a) nonlinear scattering, propagation, 
absorption and so on for the sound waves; (b) appearance of regions 
with different vacua. Certainly, the presence of such kind of 
nonlinear effects is not the direct demonstration of the PT 
but it can indicate the important role of the strong 
phonon-phonon interaction in the High-T$_c$ superconductor. 

\section{Conclusions}

The basic goal of this paper is an assumption that the quantum 
solid theory with the strong nonlinear interaction of ions in the 
lattice can be similar to the QCD in which there is the strong 
gluon-gluon interaction. Such similarity can lead to the appearance 
of nonlocal objects in the quantum solid theory like the flux tube 
in the QCD. It is possible that such a mechanism is realized in the 
High-$T_c$ superconductor by such a way that the phonons are 
confined into a nonlocal object (tube) 
located between two Cooper electrons. 
\par 
In this paper we have considered only one mechanism probably 
leading in emergence the PT. The problem on existence of 
other alternative such mechanisms was not discussed in this paper 
and remains while open. 
\par 
It is very important to note that our analysis shows that 
the Green's function method in the superconductivity theory 
is a realization of an algorithm proposed by Heisenberg 
for quantization of nonlinear spinor field. We have shown 
that Green's function method is such realization of 
discussed Heisenberg idea. We would like to 
emphasize that the nonperturbative Heisenberg quantization 
method is much more powerful than a perturbative Feynman diagram 
techniques. 
\par
Finally, we can presuppose that in the quantum field theory 
\textbf{\textit{nonlinearity can lead to nonlocality}} 
\footnote{in the case of strong nonlinearity, of course}: 
nonlinear terms $(A)^3$ and $(A)^4$ in the QCD lead to the 
flux tube and probably such a mechanism in the High-T$_c$ 
superconductivity leads to the PT. 

\begin{acknowledgments}
This work is supported by a George Forster Research Fellowship
from the Alexander von Humboldt Foundation. I would like to
thank H.-J. Schmidt for the invitation to Potsdam University 
for research.
\end{acknowledgments}

\end{document}